\documentclass[showpacs,12pt]{revtex4}
\usepackage{amssymb}
\usepackage{graphicx}
\usepackage{dcolumn}
\usepackage{bm}
\usepackage{amsmath}
\usepackage{times}
\usepackage{amsmath}
\begin{document}

\title{\textbf{Quantum Adiabatic Algorithms and Large Spin Tunnelling}}
\author{A. Boulatov}
\email{boulatov@email.arc.nasa.gov}
\author{V.N. Smelyanskiy}
\email{Vadim.N.Smelyanskiy@nasa.gov}
\date{September 5, 2003}
\address{ NASA Ames Research Center, MS 269-3, Moffett Field, CA 94035-1000}

\begin{abstract}
We provide a theoretical study of  the quantum adiabatic evolution
algorithm with different evolution paths proposed in
\cite{Farhi:paths}. The algorithm is applied to a random binary
optimization problem (a version of the 3-Satisfiability problem)
where the $n$-bit cost function is symmetric with respect to the
permutation of individual bits. The evolution paths are produced,
using the generic control Hamiltonians $H(\tau )$ that preserve
the bit symmetry of the underlying optimization problem. In the
case where the ground state of $H(0)$   coincides  with the
totally-symmetric state of an $n$-qubit system the algorithm
dynamics is completely described in terms of the motion of a
spin-$n/2$.
 We show that different control Hamiltonians can be
parameterized by a set of independent parameters that are
expansion coefficients of $H(\tau )$ in a certain universal set of
operators. Only one of these operators can be responsible for
avoiding the tunnelling in the spin-$n/2$ system during the
quantum adiabatic algorithm. We show that it is possible to select
a coefficient for this operator that guarantees a polynomial
complexity of the algorithm for all problem instances. We show
that a successful evolution path of the algorithm always
corresponds to the trajectory of a \textit{classical} spin-$n/2$
and provide a complete characterization of such paths.
\end{abstract}

 \maketitle

%\vspace{3mm}

\section{Introduction.}

Recently a novel paradigm was suggested for the design of quantum algorithms
for solving combinatorial search and optimization problems based on quantum
adiabatic evolution \cite{Farhi}. In the quantum adiabatic evolution
algorithm (QAA) a quantum state is closely following a ground state of a
specially designed slowly time-varying control Hamiltonian $H(\tau )$. At
the beginning of the algorithm the control Hamiltonian $H(0)=H_{B}$ has a
simple form with a known ground state that is easy to prepare, and at the
final moment of time it coincides with the ``problem'' Hamiltonian $H_{P}$
which ground state encodes the solution of the classical optimization
problem in question \newline
%\textbf{basis}
\begin{eqnarray}
&&H_{P}=\sum_{\mathbf{z}}E_{\mathbf{z}}|\mathbf{z}\rangle \langle \mathbf{z}|
\label{HP} \\
&&|\mathbf{z}\rangle =|z_{1}\rangle _{1}\,\otimes |z_{2}\rangle
_{2}\,\otimes \cdots \otimes |z_{n}\rangle _{n}.  \label{basis}
\end{eqnarray}
Here $E_{\mathbf{z}}$ is a cost function defined on a set of $2^{n}$ binary
strings $\mathbf{z}=\{z_{1},\ldots ,z_{n}\}$ $z_{k}=0,1$, each containing $n$
bits. The summation in (\ref{HP}) is over the $2^{n}$ states $|\mathbf{z}%
\rangle $ forming the computational basis of a quantum computer with $n$
qubits. State $|z_{k}\rangle _{k}$ of the $k$-th qubit is an eigenstate of
the Pauli matrix $\hat{\sigma}_{z}$ with eigenvalue $1-2z_{k}\pm 1$. If at
the end of the QAA the quantum state is sufficiently close to the ground
state of $H_{P}$ then the solution to the optimization problem can be
retrieved by the measurement.

It has been shown recently \cite{Vazirani:02} that the query complexity
argument that lead to the exponential lower bound for the unstructured
search \cite{Bennett} cannot be used to rule out the polynomial time
solution of NP-complete Satisfiability problem by the quantum adiabatic
evolution algorithm (QAA).

A set of examples of the 3-Satisfiability problem has been recently
constructed \cite{Farhi:annealing,Vazirani:02} to test analytically the
power of QAA. In these examples the cost function $E_{\mathbf{z}}$ depends
on a bit-string $\mathbf{z}$ with $n$ bits, $\mathbf{z}=\{z_{1},z_{2},\ldots
,z_{n}\}$, only via a Hamming weight of the string, $w_{\mathbf{z}%
}=z_{1}+z_{2}+\ldots +z_{n}$, so that $E_{\mathbf{z}}=f\left( w_{\mathbf{z}%
}\right) $ where the function $f(w)$ is in general non-monotonic and defines
a particular instance of this ``Hamming Weight Problem'' (HWP). In \cite
{Farhi:annealing,Vazirani:02} the original version of QAA \cite{Farhi} was
applied to the HWP where the control Hamiltonian is a linear interpolation
in time between the initial and final Hamiltonians.

In this case, it was shown \cite{Farhi:annealing,Vazirani:02} that the
system can be trapped during the QAA in a local minimum of the cost function
for a time that grows exponentially in the problem size $n$. It was also
shown \cite{Farhi:annealing} that an exponential delay time in the quantum
adiabatic algorithm can be interpreted in terms of the quantum-mechanical
tunnelling of an auxiliary large spin between the two intermediate states.

The above example has a significance greater than just being a particular
simplified case of a binary optimization problem with symmetrized cost.
Indeed, one can argue that it shows a generic mechanism for setting
``locality traps'' in the 3-Satisfiability problem \cite{Vazirani:talk}. But
most importantly, this example demonstrates that exponential complexity of
QAA can result from a \emph{collective phenomenon} in which transitions
between the configurations with low-lying energies can only occur by
simultaneous flipping of large clusters containing order-n bits. In spin
glasses, there is typically an exponential number of such configurations,
the so-called local ground states. A similar picture may be applicable to
random Satisfiability problems \cite{Monasson}. In some cases, these
transitions can be understood and described in terms of macroscopic quantum
tunnelling. A tunnelling of magnetization was observed in large-spin
molecular nanomagnets \cite{Wernsdorfer} and in disordered ferromagnets \cite
{Brooke:Nature}.

The paper \cite{Farhi:paths} suggests that large tunnelling barriers can be
avoided in QAA by using multiple runs of QAA with realizations of the
control Hamiltonians $H(\tau )$ sampled from a random ensemble. This
ensemble is chosen in a sufficiently simple and general form that does not
depend on the specific instance of the optimization problem. Different
Hamiltonians $H(\tau )$ correspond to different paths of the unitary
evolution that begin and end in the same initial and final states (modulus
phase factors). The complexity of QAA with different paths for the HWP was
tested numerically in \cite{Farhi:paths} using an ensemble of random 8$%
\times $8 matrices. The results indicate that the HWP may be solved in
polynomial time with finite probability.

In case when the random paths $H(\tau )$ preserve the bit-permutation
symmetry of the problem it is natural to describe the random ensemble of $%
H(\tau )$ in terms of the dynamics of a spin-$n/2$ system. This approach
allows for a general theoretical analysis of the algorithm. In the present
paper, we perform this analysis for the random version of HWP
(over-constrained 3-Satisfiability problem) by mapping the dynamics of QAA
onto the motion of a quantum particle in a 1D effective potential. This
allows us to compute the statistical weight of the successful evolution
paths in the ensemble and provide a complete characterization of such paths.

\section{Quantum Adiabatic Evolution Algorithm with different paths}

In a QAA with different paths \cite{Farhi:paths}, one specifies the
time-dependent control Hamiltonian $\tilde{H}(t)\equiv H(\tau )$
%\\{\bf H\_tot}
\begin{eqnarray}
&&H(\tau )=(1-\tau )\,H_{B}+\tau (1-\tau )\,H_{E}+\tau \,H_{P},
\label{H_tot} \\
&&\tau =\frac{t}{T}\,\in (0,1).  \notag
\end{eqnarray}
\noindent where the control parameter $\tau $ plays the role of
dimensionless time. This Hamiltonian guides the quantum evolution of the
state vector $|\psi (t)\rangle $ according to the Schr\H{o}dinger equation $%
i\hbar \,{\partial |\psi (t)\rangle \partial t}=H(\tau )|\psi (t)\rangle $
from $t=0$ to $t=T$, the \emph{run time} of the algorithm. $H_{P}$ is the
``problem'' Hamiltonian given in (\ref{HP}). $H_{B}$ and $H_{E}$ are
`driver'' Hamiltonians designed to cause the transitions between the
eigenstates of $H_{P}$.

An initial state of the system $|\psi (0)\rangle $ is prepared as a ground
state of the initial Hamiltonian $H(0)=H_{B}$. It is typically constructed
assuming $\mathit{no}$ knowledge of the solution of the classical
optimization problem and related ground state of $H_{P}$. In the simplest
case %\\{\bf HD}
\begin{equation}
H_{B}=-C\,\sum_{j=1}^{n}\sigma _{x}^{j},\quad |\psi (0)\rangle
=2^{-n/2}\sum_{\mathbf{z}}|\mathbf{z}\rangle ,  \label{HD}
\end{equation}
where $\sigma _{x}^{j}$ is a Pauli matrix for $j$-th qubit and $C>0$ is some
scaling constant. The ground state of $H_B$ has equal projections on any of
the $2^{n}$ basis states $|\mathbf{z}\rangle$ (\ref{basis}).

Consider instantaneous eigenstates $|\phi _{k}(\tau )\rangle $ of $H(\tau )$
with corresponding eigenvalues $E_{k}(\tau )$ arranged in non-decreasing
order at any value of $\tau \in (0,1)$%
%\\{\bf adiab}
\begin{equation}
H(\tau )|\phi _{k}(\tau )\rangle =\lambda _{k}(\tau )|\phi _{k}(\tau
)\rangle ,\quad k=0,1,\ldots ,2^{n}-1.  \label{adiab}
\end{equation}
Provided the value of $T$ is large enough and there is a finite gap for all $%
t\in (0,T)$ between the ground and exited state energies, $\Delta \lambda
(\tau )=\lambda _{1}(\tau )-\lambda _{0}(\tau )>0$, quantum evolution is
adiabatic and the state of the system $|\psi (t)\rangle $ stays close to an
instantaneous ground state, $|\phi _{0}(t/T)\rangle $ (up to a phase
factor). Because $H(\tau )=H_{P}$ the final state $|\psi (T)\rangle $ is
close to the ground state $|\phi _{0}(\tau =1)\rangle $ of the problem
Hamiltonian. Therefore a measurement performed on the quantum computer at $%
t=T$ will find one of the solutions of combinatorial optimization problem
with large probability. Quantum transition away from the adiabatic ground
state occurs most likely in the vicinity of the point $\tau \approx \tau _{c}
$ where the energy gap $\Delta \lambda (\tau )$ reaches its minimum
(avoided-crossing region). The probability of the transition is small
provided that \cite{adiabatic_theorem} %\\{\bf mingap}
\begin{equation}
T\gg \hbar \dot{H}_{\max }\,\Delta \lambda _{\mathrm{min}}^{-2},
\label{mingap}
\end{equation}
\noindent where
\begin{eqnarray}
\dot{H}_{\max } &=&\max_{\tau \in (0,1)}|\langle \phi _{1}(\tau )|\frac{d%
\tilde{H}}{d\tau }|\phi _{0}(\tau )\rangle |,  \notag \\
\Delta \lambda _{\mathrm{min}} &=&\min_{\tau \in (0,1)}\left[ \lambda
_{1}(\tau )-\lambda _{0}(\tau )\right] ,  \label{mingap1}
\end{eqnarray}
\noindent The r.h.s. in Eq.~(\ref{mingap}) gives an upper bound estimate for
the required runtime of the algorithm and the task is to find its asymptotic
behavior in the limit of large $n\gg 1$. The numerator in (\ref{mingap}) is
of the order of the largest eigenvalue of $dH/d\tau =H_{P}-H_{B}+(1-2\tau
)H_{E}$, which typically scales polynomially with $n$. However, $\Delta E_{%
\mathrm{min}}$ can scale down exponentially with $n$ and in such cases the
required runtime of the quantum adiabatic algorithm to find a solution grows
exponentially fast with the size of the input.

One should note that the second term in the r.h.s. of (\ref{H_tot}) is zero
at $\tau =0$ and $\tau =1$. Therefore, by using different driver
Hamiltonians $H_{E}$ one can design a family of (possibly random) adiabatic
evolution paths that start at $\tau =0$ in the same generically chosen
initial state and arrive at the ground state of $H_{P}$ at $\tau =1$. In
general, different paths will correspond to different minimum gaps $g_{%
\mathrm{min}}$ and one can introduce the distribution of minimum gaps. This
distribution can be used to compute the fraction of the adiabatic evolution
paths $f$ that arrive at the ground state of $H_{P}$ within polynomial time,
%\\{\bf poly}
\begin{equation}
T\leq c\,n^{-\alpha },\quad \alpha >0,\quad c=\mathcal{O}(1).  \label{poly}
\end{equation}
\noindent For a successfully designed family of paths the fraction $f$ is
bounded from below by a polynomial in $1/n$ which leads to the average
polynomial complexity of QAA.

\section{\label{3sat:hwp} Binary Optimization Problem with Symmetric Cost
Function}

Consider a binary optimization problem defined on a set of $n$-bit strings $%
\mathbf{z}$ with the cost function $E_{\mathbf{z}}$ in the following form:
%\\{\bf symcost}
\begin{equation}
E_{\mathbf{z}}=f\left( w_{\mathbf{z}}\right) ,\quad w_{\mathbf{z}%
}=\sum_{j=1}^{n}z_{j}.  \label{symcost}
\end{equation}
This cost is symmetric with respect to the permutation of bits, it depends
on a string $\mathbf{z}$ only through the number of unit bits in the string $%
w_{\mathbf{z}}$ (the Hamming weight). In this paper we consider the cost
function (\ref{symcost}) in the following form which is generalization of
the cost introduced in \cite{Vazirani:01,Farhi:annealing,Vazirani:02}
%\\{\bf 3sat\_gen}
\begin{eqnarray}
E_{\mathbf{z}} &=&\sum_{i_{1}<i_{2}<i_{3}}c(z_{i_{1}}+z_{i_{2}}+z_{i_{3}}),
\label{3sat_gen} \\
c(m) &=&p_{0}\delta _{m,0}+p_{1}\delta _{m,1}+p_{2}\delta _{m,2}+p_{3}\delta
_{m,3}.  \notag
\end{eqnarray}
Here the sum is over all possible 3-bit subsets of the $n$-bit string $%
\mathbf{z}$. A subset $z_{i_{1}}+z_{i_{2}}+z_{i_{3}}$ contributes to the
total cost a weight factor $p_{k}$ where $k$ is a number of units bits in
the subset. A set of weights $\{p_{k}\}$ defines an instance of this
generalized Hamming Weight Problem (HWP). One can formulate a random version
of HWP, e.g., by drawing numbers $\{p_{k}\}$ independently from a uniform
distribution defined over a certain range.

%For the purpose of convenience we use in what following a scaled
%symmetrized variable $n_{\bf z}$ instead of the Hamming weight
%function $w_{\bf z}$
%%\\{\bf nzwz}
%\begin{equation}
%n_{\bf z}=1-\frac{w_{\bf z}}{l}, \qquad l=\frac{n}{2},\label{nzwz}
%\end{equation}
%\noindent here $-1\leq n_{\bf z} \leq 1$.
In the limit of large $n\gg 1$ the cost function (\ref{3sat_gen}) takes the
following form: %corresponds to the following form of $f(w_{\bf z })$
%\\{\bf Gp}
\begin{equation}
E_{\mathbf{z}}=l^3\,G_P\left(1-\frac{w_{\mathbf{z}}}{l}\right), \qquad
G_P(q)=\sum_{k=0}^{3}\beta_{k} q^k,  \label{Gp}
\end{equation}
\noindent here $l=n/2$ and we only keep the terms of the leading order in $n$%
. The coefficients $\beta_k$ in (\ref{Gp}) are linear combinations of $p_k$
%\\{\bf beta\_1}
\begin{equation}
\beta_{k}=\frac{\xi_k}{2}\left[ p_1 +(-1)^k p_2\right]+\frac{1}{6} \binom{3}{%
k}\left[p_0+(-1)^k p_3\right].  \label{beta_1}
\end{equation}
\noindent here $\xi_k=1$ for $k=0,1$ and $\xi_k=-1$ for $k=2,3$.

The function $G_{P}(q)$ in (\ref{Gp}) is a third degree polynomial in $q$,
and the form of the function depends on the coefficients $\beta _{k}$ ($p_{k}
$). It is easy to show that there is a finite size region in the parameter
space $\{p_{k}\}$ where $G_{P}(q)$ is a non-monotonic function of $q$ that
has global and local minima on the interval $q\in (-1,1)$. Those minima are
separated by a finite barrier with width $\delta q=\mathcal{O}(1)$. The
barrier separates strings that have close values of the cost $E_{\mathbf{z}}$
but are at large Hamming distance from each other: they have $\mathcal{O}(n)$
distinct bits. This property can lead to exponentially small minimum gaps in
QAA due to the onset of low-amplitude quantum tunnelling \cite
{Farhi:annealing}.
\begin{figure}[t]
\includegraphics[width=3.3in]{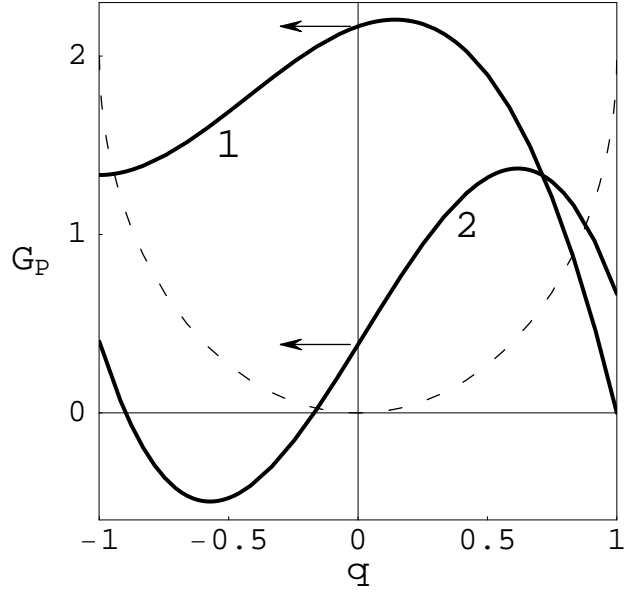}%\vspace{0.1in}
\hspace{0.2in}
\caption{Plots of the cost function (\ref{Gp}) $G_{p}$ \textit{vs} $q$ for
different choices of the weights $\{p_{k}\}$. Curve $1$ corresponds to $%
p_{0}=0,$ $p_{1}=3,$ $p_{2}=1,$ $p_{3}=1$, and the cost function $G_{P}(q)$
has a global minimum at $q=1$, corresponding to the string $\mathbf{z}$ with
the Hamming weight zero, $z_{1}=z_{2}=\ldots =z_{n}=0$. It also has a local
minimum at $q=-1$ corresponding to the bit string with Hamming weight $n$, $%
z_{1}=z_{2}=\ldots =z_{n}=1$. The curve $1$ yields the particular form of
the cost function $G_{P}(q)$ considered in \protect\cite{Farhi:annealing},
\protect\cite{Farhi:paths}. Curve $2$ corresponds to $p_{0}=0.5,$ $p_{1}=2.5,
$ $p_{2}=-2,$ $p_{3}=0.3$, it has a global minimum at $q=q^{\ast }$ inside
of the interval $(-1,1)$. This minimum corresponds to approximately $\binom{%
n\,\,}{n\,w^{\ast }}$ bit strings $\mathbf{z}$ that all have the same
Hamming weight $w_{\mathbf{z}}=w^{\ast }=n(1-q^{\ast })/2$.}
\label{fig:cost}
\end{figure}

\section{Construction of the Control Hamiltonian}

\subsection{Representation in terms of operator components of the total spin}

It is natural to consider the control Hamiltonians (\ref{H_tot}) for solving
the HWPs that are symmetric with respect to permutation of individual bits (%
\ref{basis}). In what follows, we use the normalized components of the total
spin operator $\mathbf{\hat{S}}$ for the system of $n$ individual spins-$%
\frac{1}{2}$ %\\{\bf S}
\begin{equation}
\hat{n}_{j}=\frac{1}{l}\hat{S}_{j},\qquad \hat{S}_{j}=\frac{1}{2}%
\sum_{i=1}^{n}\hat{\sigma}_{j}^{i},\qquad l=\frac{n}{2}.  \label{S}
\end{equation}
\noindent Here $\hat{S}_{j}$ are the projections of the total spin operator
on the $j$-th axis ($j=x,y,z$) and $\hat{\sigma}_{j}^{i}$ are Pauli matrices
for the $i$-th spin. For the sake of bookkeeping, in (\ref{S}) and also
throughout the paper we use ``hats'' for the spin operators, such as $\hat{S}%
_{j}$, $\hat{n}_{j}$, and some others, in order to distinguish them from
their corresponding eigenvalues ($S_{j}$ and $n_{j}$, respectively, in the
above example).

To obtain the problem Hamiltonian (\ref{HP}) we make use of the obvious
connection between the values of the Hamming weight function $w_{\mathbf{z}}$
of an $n$-bit string $\mathbf{z}$ and corresponding eigenvalues $n_{z}$ of
the spin projection operator $\hat{n}_{z}$ %\\{\bf Sz}
\begin{equation}
\hat n_{z}|\mathbf{z}\rangle =n_{z}|\mathbf{z}\rangle,\quad n_z=1-\frac{w_{%
\mathbf{z}}}{l}.  \label{Sz}
\end{equation}
\noindent Then from Eqs.~(\ref{basis}),(\ref{Gp}) and (\ref{Sz}) we obtain
%\\{\bf H\_p\_z}
\begin{equation}
H_{P}(\hat n_{z})=l^3 \,G_P\left( \hat n_{z}\right).  \label{H_p_z}
\end{equation}

We chose the driver $H_B$ in a bit-symmetric form that coincides with (\ref
{HD}) (up to a constant term)%\\{\bf HDfarhi}
\begin{equation}
H_{B}=l^3\,G_B(\hat n_x), \quad G_B(x)\equiv 2 (1-x).  \label{HDfarhi}
\end{equation}

\subsection{\label{sec:bit_sym} Bit-symmetric drivers $H_E$}

It was proposed in \cite{Farhi:paths} that $H_{E}$ can be constructed using
some generic ensemble of random matrices.
%We note that unlike (\ref{f_gen}) an  instance of the generalized
%3-SAT HWP in (\ref{f_gen}) is defined by a set of parameters
%$\{p_k\}$ that control the number and positions of its minima. The
%natural question is wether one can make a \lq\lq universal" choice
%of $H_E$ that guarantees to solve of all problem instances in
%polynomial time.
The bit-symmetric random drivers for the cost functions of the type (\ref
{3sat_gen}) can be constructed as follows \cite{Farhi:paths}. One generates
an $8\times 8$ random Hermitian matrix $A$ with zero diagonal elements and
non-diagonal elements that are independent random numbers identically
distributed in a certain interval. Matrix elements of $A_{z_{i},z_{j},z_{k}}$
can be enumerated by all possible configurations of a 3-bit string $%
\{z_{i},z_{j},z_{k}\}$. Then $H_{E}$ takes the form %\\{\bf HEA}
\begin{eqnarray}
&&H_{E}=\sum_{i<j<k}\sum_{\mathbf{z}\in
\{0,1\}^{n}}A_{z_{i},z_{j},z_{k}}\, |{\bf z}_{\bar i\,\bar j\,\bar
k}\rangle \langle
{\bf z}_{\bar i\,\bar j\,\bar k}| , \label{HEA}\\
&&|{\bf z}_{\bar i\,\bar j\,\bar k}\rangle =|z_{1}\rangle
_{1}\,\otimes\cdots \otimes|\bar{z}_{i}\rangle _{i}\,\otimes
\cdots \otimes|\bar{z}_{j}\rangle
_{j}\otimes\cdots\otimes|\bar{z}_{k}\rangle
_{k}\otimes\cdots\otimes |z_{n}\rangle _{n}. \
\end{eqnarray}
\noindent Here $|\mathbf{z}\rangle $ are computational basis
states Eq.~(\ref {HP}) corresponding to bit-strings
$\mathbf{z}=\{z_{1},\ldots ,z_{n}\}$, and string ${\bf z}_{\bar
i\,\bar j\,\bar k}$ has three of its bits flipped at the positions
$i$,$j$ and $k$  as compared to the string ${\bf z}$ (i.e.,
$\bar{z}_i=1-z_i$, etc).
 Each randomly
selected $A$ generates $H_{E}$ and therefore a random path
modification of the QAA. %It was numerically
%demonstrated in \cite{Farhi:paths} for  the specific choice of the
%weights $\{p_k\}$  in the cost function (\ref{3sat_gen}) (see
%Fig.~\ref{cost}) that a significant proportion (approximately
%$1/3$) of all random path  modifications in QAA avoids the
%macroscopic tunnelling and leads to polynomial gaps.

From the above discussion, it follows that the matrix of the operator $H_{E}$
(\ref{HEA}) is symmetric with respect to the bit permutations and therefore
it commutes with the operator of a total spin $\hat{S}^{2}$ of a system of $n
$ spins $\frac{1}{2}$. This means that $H_{E}$ acts independently in each of
the sub-spaces corresponding to certain values of the total spin $0\leq
l\leq \frac{n}{2}$ \cite{LANDAU1}. It follows from (\ref{H_p_z}) and (\ref
{HDfarhi}) that the same is true for the total control Hamiltonian (\ref
{H_tot}) %\\{\bf commut}
\begin{equation}
\left[ H(\tau ),\hat{\mathbf{S}}^{2}\right] =0,\quad \tau \in (0,1).
\label{commut}
\end{equation}
\noindent Since in our case the initial state (\ref{HD}) is a totally
symmetric combination of all states and therefore corresponds to the maximal
spin $l=\frac{n}{2}$, our system always stays in this sub-space during the
algorithm. Therefore in the analysis of the complexity of QAA one can reduce
the $2^{n}\times 2^{n}$ matrix of $H(\tau )$ to the $(2n+1)\times (2n+1)$
matrix that only involves the states with different spin projections of the
\textit{maximum} total spin $l=\frac{n}{2}$. Binary strings corresponding to
the quantum states from this subspace are distinguished from each other by
their Hamming weight only.

In Appendix \ref{matrix}, we show that in the case of real-valued symmetric
matrices $A$ and in the large-spin limit, the bit-symmetric driver $H_{E}$ (%
\ref{HEA}) can be presented as a linear combination of 6 operators expressed
in terms of the large spin operator components $\hat{n}_{x},\hat{n}_{z}$
acting in the subspace with $l=\frac{n}{2}$. Using this fact, and also Eqs.~(%
\ref{H_p_z}) and (\ref{HDfarhi}) one can write a bit-symmetric control
Hamiltonian (\ref{H_tot}) in the following form %\\{\bf H\_tot\_4}
%%\begin{widetext}
\begin{equation}
\frac{1}{l^{3}}H(\tau )\equiv G(\tau ,\hat{n}_{x},\hat{n}_{z})=(1-\tau
)\,G_{B}(\hat{n}_{x})+\tau (1-\tau )\,G_{E}\left( \hat{n}_{x},\hat{n}%
_{z}\right) +\tau \,G_{P}\left( \hat{n}_{z}\right) ,  \label{H_tot_4}
\end{equation}
\noindent %\\{\bf H\_e\_3}
\begin{equation}
G_{E}(\hat{n}_{x},\hat{n}_{z})=\gamma _{1}\hat{n}_{x}+\gamma _{2}\hat{n}%
_{x}^{2}+\gamma _{3}\hat{n}_{x}^{3}+\gamma _{4}\hat{n}_{x}\hat{n}_{z}+\gamma
_{5}\hat{n}_{x}\hat{n}_{z}^{2}+\gamma _{6}\hat{n}_{x}^{2}\hat{n}_{z},
\label{H_e_3}
\end{equation}
%%\end{widetext}
\noindent where $\left\{ \gamma _{k}\right\} $ ($k=1,\ldots ,6$)
are independent real coefficients given in Eq.~(\ref{gammas_1}).
As we show in Appendix \ref{matrix}, any random realization of the
real matrix $A$ can be mapped onto combinations of drivers
(\ref{H_e_3}) by the appropriate choice of the real coefficients
$\gamma _{k}$.

We note that $H_E$ in Eq.~(\ref{H_e_3})  does not have any terms
involving $\hat n_y$ operator. The  reason for that is that we
chose  matrix elements of $A_{z_i\,z_j\,z_k}$ (\ref{HEA}) to be
real numbers. Then matrix elements of $H_E$ in ${\bf z}$ basis are
real as well. In this case $H_E$  can only involve terms with
\textbf{even} powers of $\hat n_{y}$. In (\ref{H_e_3}) we have
used a conservation of the total spin ( see discussion above) and
substituted $n_{y}^{2}=1-n_{x}^2-n_{z}^2$.

 The form of the total Hamiltonian in (\ref{H_tot_4}) allows us to analyze the
minimum gap in QAA with different paths (\ref{H_tot}) using the
WKB analysis of the dynamics of a spin-$\frac{n}{2}$ in the large
spin limit ($n\gg 1$).

\section{Adiabatic Evolution of a Large Spin}

\subsection{WKB approximation for the large spin}

Our analysis in this section is a particular application of the WKB-type
approach commonly used for the description of quantum spin tunnelling in
magnetics \cite{CHUD1}, \cite{GARG1}, \cite{GARG2}, \cite{Brooke:Nature}.
This approach is applicable for the large spins $(l>>1)$, which is the case
of interest for us.

We choose $z$ as a quantization axis and following the standard procedure to
obtain the effective quasi-classical Hamiltonian in polar coordinates $%
\left\{ \theta ,\varphi \right\} $ with $\theta \in \left[ 0,\pi \right] $
and $\varphi \in \left[ 0,2\pi \right] $. We make use of the Villain
transformation \cite{ENZ1} %\\{\bf Villain\_1}
\begin{equation}
\hat{n}_{x}=\sqrt{1+\epsilon -\hat{n}_{z}\left( \hat{n}_{z}+\epsilon \right)
}\cos \left( \hat{\varphi}\right) ,\quad \epsilon =\frac{1}{l},
\label{Villain_1}
\end{equation}
where azimuthal angle operator $\hat{\varphi}$ satisfies the commutation
relation
\begin{equation}
\left[ \hat{\varphi},\hat{n}_{z}\right] =i\epsilon .  \label{comm1}
\end{equation}
In a change of notation we introduce a coordinate $q$ and
canonically-conjugate momentum $\hat{p}$ (cf. \cite{ENZ1}) %\\{\bf pz}
\begin{equation}
q=\hat{n}_{z},\qquad \hat{p}=-i\epsilon \frac{d}{dq}\equiv -\hat{\varphi},
\end{equation}
\noindent ($-1\leq q\leq 1$). Expanding (\ref{Villain_1}) in the large spin
limit $\epsilon \ll 1$, we obtain %\\{\bf Villain\_2}
\begin{equation}
\hat{n}_{x}=(1-q^{2})^{1/2}\cos \hat{p}+\epsilon \frac{\cos \hat{p}}{2(1+q)}+%
\mathcal{O}(\epsilon ^{2}).  \label{Villain_2}
\end{equation}
\noindent Finally, we write the scaled Hamiltonian of the system (\ref
{H_tot_4}) in terms of the new variables %\\{\bf H\_expansion}
\begin{eqnarray}
&&G(\hat{n}_{x},\hat{n}_{z},\tau )\equiv \mathcal{H}(q,\hat{p},\tau ),
\notag \\
&&\mathcal{H}(q,\hat{p},\tau )=\mathrm{H}(q,\hat{p},\tau )+\Delta \mathrm{H}%
(q,\hat{p},\tau ),  \label{H_expansion}
\end{eqnarray}
\noindent where %\\{\bf Hqp}
\begin{equation}
\mathrm{H}(q,\hat{p},\tau )=G\left( \sqrt{1-q^{2}}\cos \hat{p},q,\tau
\right) ,  \label{Hqp}
\end{equation}
\vspace{-0.2in} %\\{\bf [qp]}
\begin{equation}
\left[ q,\hat{p}\right] =i\,\epsilon ,  \label{[qp]}
\end{equation}
\noindent and $\Delta \mathrm{H}$ is a small correction %\\{\bf Hqpcorr}
\begin{equation}
\Delta \mathrm{H}(q,\hat{p},\tau )=\epsilon \,\frac{\cos \hat{p}}{2(1+q)}%
\frac{\partial G}{\partial n_{x}}+\mathcal{O}(\epsilon ^{2}),
\label{Hqpcorr}
\end{equation}
(here $\partial G/\partial n_{x}$ has the same arguments as $G$ in (\ref{Hqp}%
)).

The stationary Schr\H{o}dinger equation (\ref{adiab}) in the new basis
%\\{\bf schH}
\begin{equation}
\mathcal{H}(q,\hat{p},\tau )\Psi _{k}(q;\tau )=\lambda _{k}(\tau )\Psi
_{k}(q;\tau ),  \label{schH}
\end{equation}
\noindent can be solved in the WKB approximation with the small parameter $%
\epsilon \ll 1$ playing the role of a Plank constant. Then the wave function
$\Psi (q)$ takes the form %\\{\bf wkb}
\begin{equation}
\Psi (q)=B(q)\,\exp \left[ \frac{iA(q)}{\epsilon }\right] ,  \label{wkb}
\end{equation}
\noindent where in the leading order in $\epsilon $ the action function $A(q)
$ satisfies the following Hamilton-Jacobi equation %\\{\bf HJ}
\begin{equation}
\mathrm{H}\left( q,\frac{dA(q)}{dq},\tau \right) =\lambda .  \label{HJ}
\end{equation}
\noindent This equation describes a 1D auxiliary mechanical system with
coordinate $q$, momentum $p$, energy $\lambda $, and Hamiltonian function $%
\mathrm{H}(q,p,\tau )$. Classical orbits satisfy the Hamiltonian equations
%\\{\bf ham\_eq}
\begin{equation}
\dot{q}(t)=\mathrm{H}_{p}(q(t),p(t),\tau ),\quad \dot{p}(t)=-\mathrm{H}%
_{q}(q(t),p(t),\tau ),  \label{ham_eq}
\end{equation}
\noindent where $H_{q}$ and $H_{p}$ stand for the partial derivatives of $H$
with respect to $q$ and $p$, respectively. Stationary points of the dynamics
$(q_{\ast },p_{\ast })$ correspond to the elliptic and saddle points of the
Hamiltonian function %\\{\bf elliptic}
\begin{equation}
\mathrm{H}_{q}(q_{\ast },p_{\ast },\tau )=0,\quad
\mathrm{H}_{p}(q_{\ast },p_{\ast },\tau )=0,  \label{elliptic}
\end{equation}
\noindent Elliptic points are minima (or maxima) of $H(q,p,\tau )$ on the ($%
q,p$) plane. They satisfy the condition %\\{\bf d}
\begin{equation}
\Omega _{\ast }^{2}=H_{pp}(q_{\ast },p_{\ast },\tau
)H_{qq}(q_{\ast },p_{\ast },\tau )-H_{qp}^{2}(q_{\ast },p_{\ast
},\tau )>0.  \label{d}
\end{equation}
\noindent where $H_{pp}$ is understood as a second derivative of $H$ with
respect to $p$, etc. Saddle points correspond to $\Omega _{\ast }^{2}<0$ in (%
\ref{d}).

In the limit $\epsilon \ll 1$ the adiabatic ground state $\Psi _{0}(q;\tau )$
(\ref{schH}) is localized in the small vicinity of the fixed points $%
(q_{\ast },p_{\ast })$ corresponding to the global minimum of $\mathrm{H}%
(q,p,\tau )$ at a given value of $\tau $. To logarithmic accuracy the
WKB-asymptotic (\ref{wkb}) of the ground-state wave function is determined
by the mechanical action for the imaginary-time instanton trajectory $%
(q(t),p(t))$ emanating from the fixed point $(q_{\ast },p_{\ast })$
%\\{\bf inst}
\begin{eqnarray}
&&\hspace{-0.1in}\Psi _{0}(q;\tau )\approx \overline{\Psi }_{0}(q;q_{\ast
},p_{\ast },\tau )\propto \exp \left[ -\frac{i}{\epsilon }\int_{-i\infty
}^{0}dt\,\dot{q}(t)p(t)\right] ,  \notag \\
&&q(-i\infty )=q_{\ast },\quad p(-i\infty )=p_{\ast },\quad q(0)=q.
\label{inst}
\end{eqnarray}
\noindent Integration in (\ref{inst}) is along the imaginary axis $(-i\infty
,0)$. The instanton trajectory obeys Eq.~(\ref{ham_eq}) with the boundary
conditions given above and $t\in (-i\infty ,0)$ corresponding to the line of
integration in (\ref{inst}). The choice of the final instant, $t=0$, is
arbitrary since the instanton trajectory is degenerate with respect to a
shift of the time axis.

We note that the WKB asymptotic (\ref{inst}) decays exponentially fast as
the coordinate $q$ in (\ref{inst}) moves away from its value at the global
minimum $q_{\ast }$ into the classically inaccessible region. This
corresponds to the growth of the imaginary part of the action in (\ref{inst}%
), similar to the conventional quantum tunnelling in the potential. In the
vicinity of $(q_{\ast },p_{\ast })$ the ground-state wave function $\Psi
_{0}(q)$ takes the form similar to that of harmonic oscillator:
%\\{\bf osc}
\begin{eqnarray}
&&\Psi _{0}(q)=c\times \exp \left[ \frac{i}{\epsilon }\left( p_{\ast
}\,\delta q-\frac{H_{qp}}{2H_{pp}}\delta q^{2}\right) -\frac{m_{\ast
}\,\Omega _{\ast }^{2}\delta q^{2}}{2\epsilon }\right] ,  \notag \\
&&m^{\ast }=\frac{1}{|H_{pp}(q_{\ast },p_{\ast },\tau )|},  \label{osc}
\end{eqnarray}
\noindent here $\omega _{\ast }>0$ is defined in (\ref{d}). Similarly, the
energy spectrum in that region corresponds to the classical elliptic orbits
with oscillation frequency $\Omega _{\ast }$ %\\{\bf Omega*}
\begin{equation}
\lambda _{k}-\mathrm{H}(q_{\ast },p_{\ast },\tau )\sim \epsilon \,\Omega
_{\ast }\left( k+\frac{1}{2}\right) ,\quad k=0,1,\ldots .  \label{Omega*}
\end{equation}
\noindent We note that the frequency $\Omega _{\ast }$ depends on $\tau $
and determines the time-varying instantaneous gap between the ground and
first exited states, $\Delta \lambda =\epsilon \Omega _{\ast }(\tau )$.

\section{Local and global bifurcations during the QAA}

It can be seen from Eq.~(\ref{Hqp}) that the global minimum of $\mathrm{H}%
(q,p,\tau )$ will correspond to $p_{\ast }=\pm k\pi $ ($k=0,\pm 1,\ldots $)
provided that the following condition holds for all $n_{x}$: %\\{\bf dGdnx}
\begin{equation}
\omega _{x}\equiv -\frac{\partial G\left( n_{x},q_{\ast }(\tau ),\tau
\right) }{\partial n_{x}}\neq 0,  \label{dGdnx}
\end{equation}
where the positive and negative signs of $\omega _{x}$ correspond to even
and odd values of $k$, respectively. The value of $q_{\ast }$ in (\ref{dGdnx}%
) corresponds to the global minimum of the effective potential $U(q,\tau )$
%\\{\bf U}
\begin{equation}
U(q)=G(\sqrt{1-q^{2}},q,\tau ),\quad U(q)-U(q_{\ast })>0.  \label{U}
\end{equation}
Under the above condition the Hamiltonian function of the system near the
global minimum $(q_{\ast },0)$ exactly corresponds to that of the harmonic
oscillator with effective frequency $\Omega _{\ast }$ (\ref{d}) and mass $%
m_{\ast }$ (\ref{osc}) %\\{\bf osc1}
\begin{eqnarray}
&&\mathrm{H}(q,p,\tau )=\frac{1}{2m_{\ast }(\tau )}p^{2}+\frac{m_{\ast
}\Omega _{\ast }^{2}(\tau )(q-q_{\ast }(\tau ))^{2}}{2},  \label{osc1} \\
&&\frac{1}{m_{\ast }}=-\sqrt{1-q_{\ast }^{2}}\,\frac{\partial G\left( \sqrt{%
1-q_{\ast }^{2}},q_{\ast },\tau \right) }{\partial n_{x}},  \notag \\
&&m_{\ast }\,\Omega _{\ast }^{2}=U^{\prime \prime }(q_{\ast }).  \notag
\end{eqnarray}
\noindent In the WKB picture the ground state of the system correspond to
the particle performing zero-level oscillations near the bottom of the
slowly varying potential $U(q,\tau )$. There are two types of the
bifurcations that can destroy the above adiabatic picture:

\subsubsection{Local bifurcation}

Assume that at some instant of time $\tau =\tau _{0}$ the effective mass $%
m_{\ast }(\tau )$ goes to infinity. In the vicinity of this point the
Hamiltonian function (\ref{Hqp}) can be approximated as follows:
%\\{\bf locb}
\begin{eqnarray}
&&\hspace{-0.4cm}\mathrm{H}(q,p,\tau )=\frac{a_{0}}{4!}p^{4}-\frac{b_{0}}{2!}%
sp^{2}+\frac{c}{2!}\delta q^{2}+d_{0}s\delta q+\mathcal{O}(s^{5/2}),  \notag
\\
&&\delta q=q-q_{\ast }(\tau _{0}),\quad s=\tau -\tau _{0},  \label{locb}
\end{eqnarray}
where %\\{\bf a0b0}
\begin{eqnarray}
&&a_{0}=\frac{\partial ^{2}G}{\partial n_{x}^{2}}(1-q_{\ast }^{2}(\tau
_{0})),\quad b_{0}=\frac{\partial ^{2}G}{\partial n_{x}\partial \tau }\sqrt{%
1-q_{\ast }^{2}(\tau _{0})},  \notag \\
&&c_{0}=\frac{\partial ^{2}U}{\partial q^{2}},\quad d_{0}=\frac{\partial
^{2}U}{\partial q\partial \tau },  \label{a0b0}
\end{eqnarray}
in the above equations all functions are evaluated at the point $(q_{\ast
}(\tau _{0}),p=0)$. Equation (\ref{locb}) corresponds to $A_{3}$ bifurcation
point \cite{GILMORE1}. It can be seen from (\ref{locb}) that for $\tau >\tau
_{0}$ the single global minimum of $H(q,p,\tau )$ splits into the two minima
with nonzero momenta %\\{\bf p+-}
\begin{equation}
p_{\ast }^{\pm }(\tau )\approx \pm \left( \frac{6b_{0}(\tau -\tau _{0})}{%
a_{0}}\right) ^{1/2},\quad \tau -\tau _{0}>0.  \label{p+-}
\end{equation}
\noindent Due to the symmetry $\mathrm{H}(q,p,\tau )=\mathrm{H}(q,-p,\tau )$
the two global minima with nonzero $p_{\ast }$ will stay symmetric with
respect to the $q$-axis at later times.

It follows from (\ref{d}), (\ref{osc1})) that the linear oscillation
frequency vanishes at the bifurcation point, $\Omega _{\ast }(\tau _{0})=0$,
however the energy gap $\Delta \lambda (\tau _{0})\neq 0$. By solving the
Schr\H{o}dinger equation (\ref{adiab}) at this point in the representation
of the momentum $p$ one can find the eigenfunctions $\tilde{\Psi}_{k}(p,\tau
_{0})$ and eigenvalues $\lambda _{k}(\tau _{0})$ corresponding to a 1D
quantum system moving in a quartic potential (cf. Eq.~(\ref{locb})). This
analysis yields an estimate for the value of the gap, and the characteristic
localization range $\delta p$ for $\tilde{\Psi}_{k}(p,\tau _{0})$
\begin{equation}
\Delta \lambda \sim \epsilon ^{4/3},\quad \delta p\sim \epsilon ^{1/3}.
\label{scale}
\end{equation}

The size of the energy barrier in momentum $p$ separating the two global
minima in (\ref{locb}) grows with time for $\tau -\tau _{0}$ and this leads
to a rapid decrease of the energy gap. Sufficiently far from the bifurcation
point, $\tau -\tau _{0}\gg \epsilon ^{2/3}$, each of the global minima $%
(q_{\ast }(\tau ),p_{\ast }^{\pm }(\tau ))$ gives rise to its own WKB
asymptotic (\ref{inst}) localized at the minimum. The ground state and the
first exited state correspond to their symmetric and anti-symmetric
combinations, respectively %\\{\bf split}
\begin{eqnarray}
&&\Psi _{k}(q)=\frac{1}{\sqrt{2}}\left( \overline{\Psi }(q;q_{\ast },p_{\ast
}^{+},\tau )+(-1)^{k}\overline{\Psi }(q;q_{\ast },p_{\ast }^{-},\tau
)\right) ,  \notag \\
&&k=0,1.  \label{split}
\end{eqnarray}
\noindent For $\tau -\tau _{0}\gg \epsilon ^{2/3}$ the tunnelling splitting
of energy levels for the symmetric and antisymmetric states determines the
value of the gap $\Delta \lambda (\tau )$ and decreases exponentially fast
with $\tau -\tau _{0}$. Away from the bifurcation region, $\tau ~-~\tau
_{0}~=~\mathcal{O}(1)$, the gap scales down exponentially with $n$ (note
that $\epsilon =2/n$).

As a result of the local bifurcation, the purely adiabatic evolution in QAA
collapses. The amplitude of staying in the adiabatic ground state for $\tau
<\tau _{0}$ is nearly equally split between the states (\ref{split}) with
the two lowest eigenvalues. In general, this may reduce the probability of
finding a system in a ground state at $\tau =1$ by a factor of 2. We note
that the control Hamiltonian (\ref{H_tot_4}) is at most a cubic polynomial
in $n_{x}$, $n_{z}$, and therefore the number of local bifurcation events
during QAA is of the order of one. In the worst case they will cause the
reduction of the success probability in QAA by a constant factor.

For a given instance of the cost function (\ref{Gp}) defined by the
coefficients $\beta_k$ (or $p_k$) the onset of local bifurcations (\ref{locb}%
) depends on the choice of the driver Hamiltonian $H_E$ (\ref{H_e_3}).

There are a number of ways to select coefficients $\gamma _{k}$'s in the
driver Hamiltonian (\ref{H_e_3}) to avoid local bifurcations during QAA in a
broad range of values of the coefficients $p_{k}$. For example, to
completely suppress local bifurcations (\ref{locb}) one can keep in (\ref
{H_e_3}) only terms linear in $\hat{n}_{x}$ and set %\\{\bf avoid}
\begin{equation}
\gamma _{2}=\gamma _{3}=\gamma _{6}=0.  \label{avoid}
\end{equation}
\noindent %However, for more
%complex optimization problems the local bifurcation phenomenon can
%potentially be a root to a failure of QAA. Indeed, if the number of
%the local bifurcations events equals $K$ then the total
%probability of success in QAA will be reduced by a factor $2^K$, and
%for $K={\cal O}(n)$ this will lead to the exponential decrease in the
%success rate of the algorithm.

\subsection{Global bifurcation}

The Hamiltonian function $H=H(q,p,\tau )$ defines a 3D surface over a 2D
plane $(q,p)$ and the shape of this surface varies with time $\tau $. We
consider global bifurcations of this surface where the energies of its two
minima cross each other at some instant of time $\tau =\tau _{0}$ while the
distance between the minima on the $(q,p)$ plane remains finite at the
crossing point. For $\tau >\tau _{0}$ the minima exchange their roles:
global minimum becomes local and vise versa. Before and after the
intersection in the energy space the two minima are uniquely identified with
the ground and first exited states of the system's Hamiltonian (\ref{Hqp}).
The corresponding wave functions $\Psi _{0,1}(q)$ are well approximated by
their asymptotic expressions(\ref{inst}),(\ref{osc}).

The small vicinity of the global bifurcation point can be described within
the standard 2-level avoiding-crossing picture. There $\Psi_{0,1}(q)$ are
given by symmetric and antisymmetric superpositions of the WKB-asymptotic
corresponding to intersecting minima. The value of the gap changes with time
as $\sqrt{c^2(\tau-\tau_0)^2+\Delta \lambda_{\min}^{2}}$ where $c$ is some
constant and the minimum gap is determined by the overlap of the WKB
asymptotic. To logarithmic accuracy it is given by the imaginary part of the
mechanical action (\ref{inst}) along the instanton trajectory connecting the
two minima
\begin{equation}
-\epsilon \log \Delta \lambda_{\min}= \left |\mathrm{Im}\int_{-i\infty}^{i%
\infty}dt\, \dot q(t)p(t) \right|
\end{equation}
\begin{equation}
\lim_{\tau\rightarrow \pm i\infty} q(\tau)=q_{*}^{1,2},
\lim_{\tau\rightarrow \pm i\infty} p(\tau)=p_{*}^{1,2}
\end{equation}
\noindent Here $q_{*}^{k}, p_{*}^{k}$ are coordinates of the two minima; $%
H(q_{*}^{1}, p_{*}^{1}, \tau_0)=H(q_{*}^{2},p_{*}^{2},\tau_0)$, and the
instanton trajectory obeys the Eqs.~(\ref{ham_eq}). The analytical
expression for the minimum gap was studied in \cite{Farhi:annealing}, \cite
{Vazirani:01} for the case $H_E=0$, using a simplified version of the
Hamming Weight problem (\ref{3sat_gen}). Below we identify certain
geometrical properties of the global bifurcations in the case $H_E=0$ that
will be used later in the selection of the drivers $H_E$ for the successful
QAA.

\begin{center}
\begin{figure}[th]
\hspace{0.2in} \includegraphics[width=3.3in]{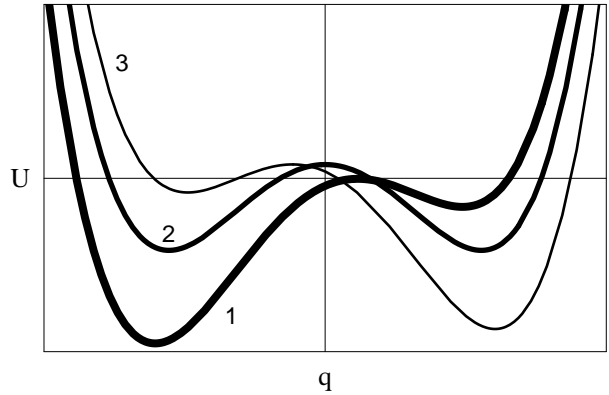}
\caption{The global bifurcation mechanism: the effective potential profiles $%
U\left( q,\protect\tau\right)$ \textit{vs} $q$ for $\protect\tau <\protect%
\tau _{0}$, $\protect\tau =\protect\tau _{0}$ and $\protect\tau >\protect\tau
_{0}$ are represented by the curves 1,2, and 3, respectively.}
\label{fig:U}
\end{figure}
\end{center}

\subsubsection{The case $H_E=0$}

In the case $\gamma _{j}\equiv 0$ $(j=1,...6)$, the Hamiltonian has a
minimum at $p_{\ast }(\tau )=\pi k$ and the value of $q_{\ast }(\tau )$
corresponds to the global minimum of the effective potential $U(q,\tau )$ (%
\ref{U}). We use Eq.~(\ref{H_tot_4}) and also the condition $U^{\prime
}(q_{\ast })=0$ to obtain the following equation for $q_{\ast }(\tau )$
%\\{\bf dq*dt}
\begin{equation}
\frac{dq_{\ast }(\tau )}{d\tau }=-\frac{G_{P}^{\prime }(q_{\ast }(\tau ))}{%
(1-\tau )\,U^{\prime \prime }(q_{\ast }(\tau ),\tau )}.  \label{dq*dt}
\end{equation}
\noindent This equation holds until the global bifurcation point at $\tau
=\tau _{0}$ where $q_{\ast }(\tau )$ changes discontinuously in time (see
Fig.~ 2). At the minimum of the potential $U^{\prime \prime }(q_{\ast })>0$
and therefore the direction of the motion of $q_{\ast }(\tau )$ entirely
depends on the direction of the ``force'', $-G_{P}^{\prime }(q_{\ast })$. At
$\tau =0$ the potential $U(q,0)$ has a unique minimum at the point $%
q=q_{\ast }(0)=0$. It is clear that with this initial condition equation (%
\ref{dq*dt}) can lead to a ``wrong'' minimum of $G_{P}(q)$ that lies above
the global minimum, and such cases will give rise to a global bifurcation.
This effect is illustrated in Fig.\ref{fig:cost} where the two different
cost functions correspond to the same direction of motion for $q_{\ast
}(\tau )$. The value of $q_{\ast }(\tau )$ may either smoothly approach the
global minimum of the cost (curve 2), or move toward a ''wrong'' local
minimum (curve 1), leading to the\ global bifurcation and exponentially
small gap in QAA. Adding $H_{E}$ to the control Hamiltonian can invert the
direction of motion of $q_{\ast }(\tau )$ toward the global minimum of $%
G_{P}(q)$. This can be seen from the fact the Eq.(\ref{dq*dt}) in presence
of $H_{E}$ possesses the additional term
\begin{equation}
-\frac{(1-\tau )}{U^{\prime \prime }(q_{\ast },\tau )}\frac{\partial G_{E}(%
\sqrt{1-q_{\ast }^{2}},q_{\ast })}{\partial q_{\ast }},  \label{avoid_ad}
\end{equation}
\noindent (here we drop for sake of brevity the argument $\tau $ in $q_{\ast
}(\tau )$). Clearly, the successful $G_{E}$ should \textit{not} possess
reflection symmetry with respect to $n_{z}$. Therefore we should only select
the terms in (\ref{H_e_3}) that contain odd powers of $n_{z}$. Taking into
account (\ref{avoid}) we arrive at the following form of the driver
Hamiltonian %\\{\bf gamma4}
\begin{equation}
G_{E}(\hat{n}_{x},\hat{n}_{z})=\gamma _{4}n_{x}\,n_{z}.  \label{gamma4}
\end{equation}
\noindent This driver can remove the potential barrier between the two
competing global minima of $U(q,\tau )$ by shifting the original minimum at $%
\tau =0$ towards the true global minimum of the cost function $G_{P}(q)$
(cf. Fig.~\ref{fig:cost}). In the classical picture (\ref{Hqp}) the driver (%
\ref{gamma4}) corresponds to an external field parallel to $z$-axis which
can destroy the tunnelling barrier along this direction. The mechanism of
such tunnelling avoidance is similar to the one considered in \cite
{Farhi:paths}, where the external field generated by the driver (\ref{gamma4}%
) compensates the effective field due to the linear term proportional to the
coefficient $\beta _{1}$ in the problem Hamiltonian (\ref{Gp}).

\subsection{\label{sec:bif_tun} Bifurcation transition to the tunnelling regime}

In general, one can expect that a complete suppression of the tunnelling
barrier at all values of $\tau $ requires a certain magnitude (and sign) of
the coefficient $\gamma _{4}$ depending on the choice of the coefficients $%
\beta _{k}$ in the cost function $G_{P}(q)$.

The transition to the tunnelling regime can be described as an $A_{3}$
bifurcation point, illustrated in Fig.2. The effective potential $U$ changes
parametrically with $\tau ,\gamma _{4}$ and $\left\{ \beta _{k}\right\} $.
Near the bifurcation point $(\tau _{c},\gamma _{4c},q_{c})$ ,the potential
has the form $U=a\,\delta q^{4}+b\,\delta \gamma \delta q^{2}+c\,\delta
q\delta \tau $ where $\delta \tau ,\delta q,\delta \gamma $ are deviations
from the bifurcation point in $\tau ,q$ and $\gamma _{4}$, respectively. The
corresponding conditions for the $A_{3}$ bifurcation point are:
\begin{equation}
\frac{\partial U}{\partial q}=\frac{\partial ^{2}U}{\partial q^{2}}=\frac{%
\partial ^{3}U}{\partial q^{3}}=0.  \label{condA3}
\end{equation}
\noindent Taking into account (\ref{U}) and (\ref{H_tot_4}),(\ref{H_e_3}),(%
\ref{gamma4}), the above equation yields %\\ {\bf bifurcation\_1}
\begin{eqnarray}
\tau _{c}\left( 1-\beta _{2}\right)  &=&1+\frac{2}{3}\frac{\tau
_{c}^{2}\left( 1-\tau _{c}\right) \left( \beta _{1}+\beta _{3}\right) ^{2}}{%
\left[ 2-\tau _{c}\left( 2-\beta _{2}\right) \right] ^{2}},
\label{bifurcation_1} \\
\gamma _{4c}\left( 1-\tau _{c}\right)  &=&\frac{\left( 1-\tau _{c}\right)
\left( 3\beta _{1}+\beta _{3}\right) +\tau _{c}\beta _{2}\beta _{3}}{\tau
_{c}\beta _{2}-2\left( 1-\tau _{c}\right) },  \notag
\end{eqnarray}
These equations should be solved for $\gamma _{4c}$ and $\tau _{c}$ for the
given set of the coefficients $\beta _{k}$. The bifurcation is avoided when
%\\{\bf avoid\_gamma4}
\begin{equation}
|\gamma _{4}|>|\gamma _{4c}|.  \label{avoid_gamma4}
\end{equation}
\noindent For example, in the particular case of the HWP (\ref{3sat_gen})
considered in \cite{Farhi:annealing}, \cite{Farhi:paths}, we have
%\\ {\bf avoid\_farhi}
\begin{eqnarray}
&&\beta _{1}=1/2,\quad \beta _{2}=-3/2,\quad \beta _{3}=-7/6,
\label{avoid_farhi} \\
&&\tau _{c}\approx 0.44,\quad \gamma _{4c}\approx -0.95.  \notag
\end{eqnarray}
\noindent In this case the example of the driver Hamiltonian $H_{E}$ that
allows to avoidance of tunnelling in QAA was given in \cite{Farhi:paths}
where the value of $\gamma _{4}=-8$ was used. According to (\ref{avoid_farhi}%
) this value is way below the critical value $\gamma _{4c}$.

\subsubsection{Numerical Simulations of the bifurcation boundary}

We performed numerical simulations with the effective potential (\ref{U})
checking for the onset of tunnelling for all $p_{k}\in \lbrack 0;3]$, $%
k=1,...4$. The numeric simulations confirm that the situation discussed
above is typical for the general HWP, implying that (\ref{gamma4}) is the
only driver term that can be fundamentally responsible for the tunnelling
avoidance in a general case, if the coefficient $\gamma _{4}$ is defined
appropriately. In particular, one of the two drivers (\ref{gamma4}) with
%\\{\bf avoid3}
\begin{eqnarray}
&&\gamma _{4}\geq \gamma _{c}=4.9\quad \mathrm{or}\quad \gamma _{4}\leq
-\gamma _{c}=-4.9  \notag \\
&&p_{k}\in \lbrack 0;3]\quad k=0,1,2,3,  \label{avoid3}
\end{eqnarray}
\noindent always suppresses tunnelling in the QAA.
\begin{center}
\begin{figure}[th]
\hspace{0.2in} \includegraphics[width=3.3in]{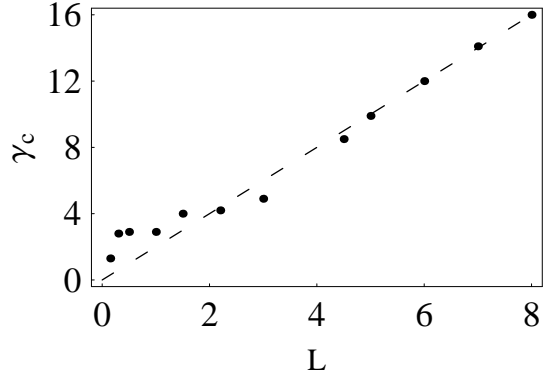}
\caption{The  critical value $\gamma_c$ {\it vs} domain size $L$.}
\label{fig:gammac}
\end{figure}
\end{center}
\noindent
We also solve the Eqs.~(\ref{bifurcation_1}) numerically for coefficients $%
p_{k}$ taking values on a dense grid of points in the cube $p_{k}\in \lbrack
0;L]$ ($k$=0,1,2,3). For each size of the cube $L$ we select the point with
largest value of $\gamma _{4c}$ denoted below as $\gamma _{c}\equiv \gamma
_{c}\left( L\right) $. The results are presented in Fig.3.
 The critical value $\gamma _{c}$ is monotonically increasing in $L$%
, and the dependence is close to linear for sufficiently large $L$, but it
is non-linear in the range $0\lesssim L\lesssim 3$. It can be inferred from
Eq.~(\ref{bifurcation_1}) that a nonlinear dependence of $\gamma _{c}$ on
the scale $L$ is due to the fact that the critical time $\tau _{c}$ also
depends on $L$.

The linear dependence of $\gamma _{c}\left( L\right) $ for large $L$ has a
simple intuition. According to (\ref{Gp}) and (\ref{beta_1}), the magnitude
of $G_{P}(q)$ is proportional to $L$. According to Eq.(\ref{dq*dt}), the
maximal magnitude of the coefficient $\beta _{1}$ presents a ''force'' that
can possibly move a system into the local minimum at small $\tau $ . From (%
\ref{beta_1}), we conclude that $\left| \beta _{1}\right| _{\max
}=\max_{p_{k}\in \lbrack -L,L]}\beta _{1}=2L$. In the limit of large $L$,
the role of the driver $G_{B}$ in (\ref{bifurcation_1}) becomes unimportant.
Therefore, the only competing terms are the driver $G_{E}$ and the problem
Hamiltonian $G_{P}$. The term (\ref{avoid_ad}) generated by $H_{E}$
compensates (\ref{avoid_ad})\ the ''force'' $\beta _{1}$ when $\left| \gamma
_{4}\right| \geq \left| \beta _{1}\right| _{\max }$, and therefore in this
limit we have
\begin{equation}
\gamma _{c}\left( L\right) =\max_{p_{k}\in \lbrack -L,L]}\gamma _{4c}\approx
\left| \beta _{1}\right| _{\max }=2L.  \label{largeL}
\end{equation}

One should note that among the effective potentials generated by choosing
different $\left\{ p_{k}\right\} $, there are two subsets that can be mapped
onto each other by means of the mirror reflection about the $q$-axis, $%
U(q,\tau )\rightarrow -U(q,t)$. We note that the same driver $H_{E}$ can not
simultaneously suppress tunnelling barriers in each of the two mutually
symmetric potentials: if the tunnelling barriers are not suppressed with $%
\gamma _{c}$, they will be suppressed with $-\gamma _{c}$, and vice versa.
This gives a simple intuition for the tunnelling barrier suppression
boundary (\ref{avoid_gamma4}).

Finally we conclude, that it is possible to indicate the range of value of $%
|\gamma _{4}|$ such that the driver Hamiltonian $H_{E}=l^{3}\,\gamma _{4}%
\hat{n}_{x}\,\hat{n}_{z}$ will play the role of a \textbf{universal} driver
that guarantees polynomial performance of the QAA for all instances of the
generalized Hamming weight problem (\ref{3sat_gen}) provisory to the
mirror-reflection symmetry in the possible choice of the cost functions and
the common normalization factor $L$.

\section{Probability of success of the QAA with random paths}

Using the analysis from the previous section one can estimate the
probability of success for the QAA with random paths proposed in \cite
{Farhi:paths}. In that algorithm, the ensemble of random drivers $H_{E}$ was
generated using random 3$\times $3 matrices $A_{z_{i}\,z_{j}\,z_{k}}$ (\ref
{HEA}). It is shown in Appendix~\ref{matrix} that for the bit-symmetric
optimization problem (\ref{3sat_gen}) the above ensemble is identical to the
ensemble of independent uniformly-distributed random coefficients $\gamma
_{k}$ (k=$1-6$) that appears in the large-spin representation of the driver $%
H_{E}$ (\ref{H_e_3}). Then for any instance of optimization problem in (\ref
{3sat_gen}) defined by the set of the coefficients $\{p_{k}\}$ one should
compute the fraction $\emph{f}\,$ of the domain of the coefficients $%
\{\gamma _{k}\}$ where the following conditions are satisfied:

\begin{enumerate}
\item[(i).]  Condition for the nonzero effective mass (\ref{dGdnx}).

\item[(ii).]  The condition (\ref{avoid_gamma4}) for the complete avoidance
of the tunnelling barriers in combination with Eq.(\ref{bifurcation_1}) for
the bifurcation boundary.
\end{enumerate}

\noindent Here we compute the fraction $f$ for the particular instance of
the optimization problem (\ref{3sat_gen}) considered in \cite
{Vazirani:01,Farhi:annealing,Farhi:paths}. In this case Eq.~(\ref{dGdnx})
takes the form
\begin{eqnarray}
\frac{\omega _{x}}{1-\tau } &=&-2+\tau \left[ \gamma _{1}+2\gamma
_{2}n_{x}+3\gamma _{3}n_{x}^{2}+\gamma _{4}q\right.   \notag \\
&&\left. +\gamma _{5}q^{2}+2\gamma _{6}n_{x}q\right] \neq 0,
\label{effmass_2}
\end{eqnarray}
where $n_{x}=\pm \sqrt{1-q_{\ast }^{2}}$ and $q_{\ast }$ provides global
minimum of $U(q,\tau )$ (\ref{U}). The effective mass is non-zero if $\omega
_{x}\neq 0$, and (\ref{effmass_2}) yields an estimate on the range of $%
\left\{ \gamma _{k}\right\} $ as %\\{\bf range}
\begin{equation}
\left| \gamma _{2}\right| +\left| \gamma _{6}\right| \leq 1+1/2\left( \left|
\gamma _{1}\right| +3\left| \gamma _{3}\right| +\left| \gamma _{4}\right|
+\left| \gamma _{5}\right| \right) .  \label{range}
\end{equation}
\noindent Following \cite{Farhi:paths} we assume that the non-diagonal
matrix elements $A_{z_{i}\,z_{j}\,z_{k}}$ are distributed in the interval $%
\left[ -3,3\right] $. Making use of (\ref{gammas_1}), we obtain%
%\\{\bf ineq}
\begin{equation}
\left| \gamma _{2}\right| +\left| \gamma _{6}\right| \leq 16,\quad \left|
\gamma _{1}\right| +3\left| \gamma _{3}\right| +\left| \gamma _{4}\right|
+\left| \gamma _{5}\right| \leq 50.  \label{ineq}
\end{equation}
\noindent Therefore, the probability that inequality (\ref{range}) is
satisfied is estimated as 1-15$^{2}/\left( \mathrm{50}\times \mathrm{16}%
\right) \approx $ 0.71875. On the other hand, the values of $\gamma _{4}$ in
(\ref{gammas_1}) belong to the range, $-12\leq \gamma _{4}\leq 12$. Using
the value of $\gamma _{c}\approx $ -0.95 given in (\ref{avoid_farhi}) we
estimate the probability of $\gamma _{4}\leq -\gamma _{c}$ to be
approximately equal to $\approx $ 0.46. Making an approximation that the
cases when the effective mass is non-zero are statistically independent from
the cases when $\gamma _{4}\leq -\gamma _{c}$, we obtain the total
probability of success as $P_{tot}\approx $0.46 $\times $ 0.71875$%
=0.334\approx $ 1/3, which is in qualitative agreement with the numerical
results of \cite{Farhi:paths}. This estimate can be generalized to the case
when the matrix elements $A_{z_{i}\,z_{j}\,z_{k}}$ are distributed in the
interval $\left[ -L,L\right] $ for sufficiently large $L>3$. In this case,
the probability that (\ref{range}) is satisfied remains the same, $\approx $
0.71875, while the probability that $\gamma _{4}\leq -\gamma _{c}$ is
estimated as $\left( 4L-\left| \gamma _{c}\right| \right) /8L$. With the
assumption of statistical independence, the total probability of success is $%
P_{tot}\approx 0.718\times \left( 4L-0.95\right) /8L$, and in the limit of
large $L>>1$ we have $P_{tot}\approx 0.359$ which exceeds slightly the value
for $L=3$.

\section{Polynomial QAA and Classical Dynamics of Large Spin}

In absence of tunnelling, the dynamics of the large spin can be
characterized by classical equations of motion for the spin projections
treated as c-numbers in the form \cite{CHUD1}
\begin{equation}
\frac{d\overrightarrow{S}}{dt}=\left[ \overrightarrow{\omega },%
\overrightarrow{S}\right] ,  \label{Heis_dyn_1}
\end{equation}
with

\begin{equation}
\overrightarrow{\omega }=\frac{\partial H}{\partial \overrightarrow{S}}%
=\left\{ \frac{\partial H}{\partial S_{x}},\ \frac{\partial H}{\partial S_{y}%
},\ \frac{\partial H}{\partial S_{z}}\right\} .  \label{Heis_dyn_2}
\end{equation}
In coordinate form and in terms of the dimensionless spin projections, this
yields

\begin{eqnarray}
\frac{dn_{x}}{dt} &=&-\omega _{z}n_{y},  \notag \\
\frac{dn_{y}}{dt} &=&\omega _{z}n_{x}-\omega _{x}n_{z},  \label{Heis_3} \\
\frac{dn_{z}}{dt} &=&\omega _{x}n_{y},  \notag
\end{eqnarray}
where we took into account that since $H$ does not contain the $S_{y}$
component, $\overrightarrow{\omega }=\frac{\partial H}{\partial
\overrightarrow{S}}=\left\{ \omega _{x},\ 0,\ \omega _{z}\right\} $. In the
case when the ''effective magnetic field'' $\overrightarrow{\omega }$ does
not explicitly depend on time, the system (\ref{Heis_dyn_1}), (\ref
{Heis_dyn_2}) has two independent integrals of motion

\begin{eqnarray}
\overrightarrow{S}^{2} &=&S_{x}^{2}+S_{y}^{2}+S_{z}^{2},  \label{int2} \\
J &=&\frac{1}{\omega }\left( \overrightarrow{\omega },\overrightarrow{S}%
\right) =\frac{\omega _{x}S_{x}+\omega _{z}S_{z}}{\sqrt{\omega
_{x}^{2}+\omega _{z}^{2}}}.  \notag
\end{eqnarray}
The first (\ref{int2}) reflects the conservation of total spin and also
holds for an arbitrary time-dependent field $\overrightarrow{\omega }=\frac{%
\partial H}{\partial \overrightarrow{S}}$, whereas the second integral
corresponds to the adiabatic invariant of the system (\ref{Heis_dyn_1}), (%
\ref{Heis_dyn_2}). Since in our case $\overrightarrow{\omega }=\frac{%
\partial H}{\partial \overrightarrow{S}}$ is parametrically time-dependent,
the adiabatic invariant is conserved approximately for sufficiently slow
parametric evolution. Note that the adiabatic solutions always play the role
of ''envelope solutions''. This means that on average, the spin closely
follows the adiabatic solution, but there are fast oscillatory-type motions
superimposed on the slow adiabatic evolution. Basically, the adiabatic
approximation in the classical case is applicable when the ''slow'' motion
is much slower than the fast oscillatory motion. This exactly corresponds to
the adiabatic evolution of the spin system in the quantum case \cite{LANDAU1}%
.

Making use of (\ref{int2}) and taking into account that at the instant $\tau
=0$, the total spin was parallel to the $x$-axis, we obtain $J/l=\frac{%
\omega _{x}n_{x}+\omega _{z}n_{z}}{\sqrt{\omega _{x}^{2}+\omega _{z}^{2}}}=1$%
, or

\begin{equation}
\overrightarrow{n}=\frac{\overrightarrow{\omega }}{\omega },
\label{adiabatic_1}
\end{equation}
implying that the total spin is always parallel the effective magnetic field
$\overrightarrow{\omega }$. Therefore, the adiabatic evolution of the large
spin can be simply described as the situation when the spin follows the
effective field (on average).

We note that at this level, there is a direct correspondence between the
adiabatic classical solution and the quasiclassical wave functions of the
large spin parallel to $\overrightarrow{n}$. From (\ref{adiabatic_1}), it
follows that this direction can be identified with the effective magnetic
field $\overrightarrow{\omega }=\frac{\partial H}{\partial \overrightarrow{S}%
}$. This justifies the ''variational'' approach introduced in \cite
{Farhi:annealing,Farhi:paths}, identifying the variational wave functions
with the adiabatic ground states along the evolution paths when the total
spin is parallel to $\overrightarrow{\Omega }$. Therefore, one can observe
that in the absence of tunnelling, the general HWP is solved essentially by
the classical paths of the QAA.

\section{Conclusions}

We apply the quantum adiabatic evolution algorithms with different paths
\cite{Farhi:paths} to the generalized Hamming Weight Problem that
corresponds to the specific case of the random Satisfiability problem
defined in (\ref{3sat_gen}). We show that any random evolution path produced
by this algorithm for the HWP can be obtained by using 6 specific
deterministic basis operators with random weights and therefore is
parameterized by 6 independent random numbers. Therefore, the approach to
QAA with different paths can still be reduced to the large spin dynamics for
the HWP. We show that only one of these ''generators'' can be a
''universal'' driver fundamentally responsible for tunnelling suppression
for arbitrary HWP and therefore the problem of constructing such a universal
driver reduces to the definition of its weight $\gamma _{4}$. Due to the
possible reflection symmetry of the cost function, any particular case of
the general HWP can be solved with one of the two values of the weight with $%
|\gamma _{4}|>\gamma _{4c}$, that is by applying one of the two universal
path modifications. We analyze the nature of the wave functions along the
successful paths and show that it is quasiclassical and corresponds to the
dynamics of a large classical spin. Therefore, we show that the general HWP
is solved by completely classical paths of the QAA and present a complete
characterization of these paths.

We analyzed in details the types of bifurcations of the effective
Hamiltonian function $H(q,p)$ that lead to the collapse of the adiabatic
evolution. The global bifurcations correspond to the onset of tunnelling in
QAA and lead to the failure of the algorithm. In contrast, the local
bifurcations while still corresponding to exponentially small minimum gap
only lead to the decrease of the probability of success by a factor of 2.
Since in a given problem function $H(q,p)$ is a low degree polynomial in its
arguments there are only a few local bifurcations possible. However, the
phenomenon of local bifurcations may become important for more difficult
random optimization problems. Assuming the number of such bifurcations $M$
is large the probability of success is reduced by a factor of $2^{-M}$. For $%
M$ that scales up with $n$ that would lead to the failure of the algorithm.

%We finally note that in our analysis we chose the elements of the
%Hermitian matrix $A_{z_i\,z_j\,z_k}$ to be real numbers. The
%generalization on the case of complex numbers $A_{z_i\,z_j\,z_k}$
%will lead to the additional terms in the driver Hamiltonian $H_E$
%(\ref{H_e_3}), involving odd powers$\hat n_y$ operator. In
%this case the universal driver term (\ref{gamma4}) will be
%modified as follows: $\gamma_4 \hat n_x \hat n_z\rightarrow
%\gamma_{4}(\cos\theta \,\hat n_x+\sin\theta\,\hat n_y)\hat n_z$.
%This corresponds to the rotation of the transverse component of
%the effective magnetic field (\ref{Heis_dyn_2}) in the $xy$ plane.
%We note however, that the form of the effective potential $U(q)$
%(\ref{U}) depends on the magnitude of the transverse field,
%$\sqrt{1-n_{z}^2}$, but not the orientation. Therefore the
%condition for the avoidance of tunnelling considered in
%Sec.~\ref{sec:bif_tun} will not change.

%The method described in the paper, suggests an interesting extension to the
%case of the K-SAT HWP in the limit of large $K$. In the large-spin limit,
%the effective potential can be described as a white noise with certain
%intensity and the well-known methods (optimal fluctuation approach) are
%applicable. This leads to the complete characterization of the minimal gap
%statistics. In particular, there always exists a range of parameters where
%the gap and therefore the run-time of QAA are polynomial in $n$ in the large-%
%$n$ limit.

%\acknowledgments
\vspace{0.2in}

\section{Acknowledgments}

We want to thank Edward Farhi (MIT) for useful discussion. This
work was supported in part by the National Security Agency (NSA)
and Advanced Research and Development Activity (ARDA) under Army
Research Office (ARO) contract number XXXXXX-XX-X-XXXX, we also
want to acknowledge the support of NASA IS Revolutionary Computing
Algorithms program (project No: 749-40).

\appendix

\section{\label{matrix} Spin operator representation of matrix $A$}

The random real symmetric $8\times 8$ random matrix $A$ introduced in (\ref
{HEA}), describes the ''transitions'' between each of the $2^{3}=8$ states
for each clause involving $3$ bits \cite{Farhi:paths}. This matrix has $%
\left( 8\times 8-8\right) /2=28$ independent matrix elements and can be
presented in the form
\begin{equation}
A=A^{\left( 1\right) }+A^{\left( 2\right) }+A^{\left( 3\right) },
\label{A_c_1}
\end{equation}
where $A^{\left( 1\right) },A^{\left( 2\right) },A^{\left( 3\right) }$
correspond to the transitions involving one, two and three bits,
respectively. For each realization, we have %\begin{widetext}
\begin{eqnarray}
A^{\left( 1\right) } &=&a_{\alpha }\ \sigma _{\alpha }^{x}\ \frac{1}{4}%
\sum_{s,s^{\prime }=\pm 1}b_{ss^{\prime }}\left( 1+s\ \sigma _{\beta
}^{z}\right) \left( 1+s\ \sigma _{\gamma }^{z}\right) ,  \label{A_c_2} \\
A^{\left( 2\right) } &=&a_{\alpha \beta }\ \left( \sigma _{\alpha
}^{+}\sigma _{\beta }^{+}+\sigma _{\alpha }^{-}\sigma _{\beta }^{-}\right) \
\frac{1}{2}\sum_{s=\pm 1}b_{s}\left( 1+s\ \sigma _{\gamma }^{z}\right)
\notag \\
&&\hspace{2.4in} +\,\widetilde{a}_{\alpha \beta }\ \left( \sigma _{\alpha
}^{+}\sigma _{\beta }^{-}+\sigma _{\alpha }^{-}\sigma _{\beta }^{+}\right) \
\frac{1}{2}\sum_{s=\pm 1}\widetilde{b}_{s}\left( 1+s\ \sigma _{\gamma
}^{z}\right) ,  \notag \\
A^{\left( 3\right) } &=&B\left( \sigma _{1}^{+}\sigma _{2}^{+}\sigma
_{3}^{+}+\sigma _{1}^{-}\sigma _{2}^{-}\sigma _{3}^{-}\right) +C\left(
\sigma _{1}^{+}\sigma _{2}^{+}\sigma _{3}^{-}+\sigma _{1}^{-}\sigma
_{2}^{-}\sigma _{3}^{+}\right)  \notag \\
&& \hspace{1.67in}+ D\left( \sigma _{1}^{-}\sigma _{2}^{+}\sigma
_{3}^{+}+\sigma _{1}^{+}\sigma _{2}^{-}\sigma _{3}^{-}\right) +E\left(
\sigma _{1}^{+}\sigma _{2}^{-}\sigma _{3}^{+}+\sigma _{1}^{-}\sigma
_{2}^{+}\sigma _{3}^{-}\right) ,  \notag
\end{eqnarray}
%%\end{widetext}
\noindent \noindent where
\begin{equation}
a_{\alpha },b_{ss^{\prime }},a_{\alpha \beta },b_{s},\widetilde{a}_{\alpha
\beta },\widetilde{b}_{s},B_{\alpha \beta \gamma },C_{\alpha \beta \gamma
},D_{\alpha \beta \gamma },E_{\alpha \beta \gamma }
\end{equation}
\noindent are the real coefficients. The indices $\left( \alpha ,\beta
,\gamma \right) \in \left\{ 1,2,3\right\} $ label the bits, $\sigma _{\alpha
}^{\pm },$ $\sigma _{\alpha }^{x}$ and $\sigma _{\alpha }^{z}$ are the Pauli
sigma-matrices of raising/lowering, $x$-projection and $z$-projection,
respectively and $s=\pm 1$ is a spin projection variable. Note that the
operator $\frac{1}{2}\left( 1+s\ \sigma _{k}^{z}\right) $ is a projector
onto the spin state $s$ for the bit $k$. Clearly, the number of independent
parameters in (\ref{A_c_2}) is $3\times 4+6\times 2+4=28$, where the three
terms of the sum correspond to $A^{\left( 1\right) },A^{\left( 2\right) }$
and $A^{\left( 3\right) }$, respectively. Note that this number of
parameters equals the number of independent matrix elements of $A$ estimated
above. In the matrix form, the representation (\ref{A_c_2}) yields
%%\begin{widetext}
\begin{equation}
A_{C}=\left[
\begin{array}{cccccccc}
0 & a_{3}b_{++} & a_{2}b_{++} & \widetilde{a}_{23}\widetilde{b}_{+} &
a_{1}b_{++} & \widetilde{a}_{13}\widetilde{b}_{+} & \widetilde{a}_{12}%
\widetilde{b}_{+} & B \\
& 0 & \widetilde{a}_{23}\widetilde{b}_{+} & a_{2}b_{+-} & \widetilde{a}_{13}%
\widetilde{b}_{+} & a_{1}b_{+-} & C & a_{12}b_{-} \\
&  & 0 & a_{3}b_{+-} & \widetilde{a}_{12}\widetilde{b}_{+} & D &
a_{1}b_{-+}
& a_{13}b_{-} \\
&  &  & 0 & E & \widetilde{a}_{12}\widetilde{b}_{-} & \widetilde{a}_{13}%
\widetilde{b}_{-} & a_{1}b_{--} \\
&  &  &  & 0 & a_{3}b_{-+} & a_{2}b_{-+} & a_{23}b_{-} \\
&  &  & &  & 0 & \widetilde{a}_{23}\widetilde{b}_{-} & a_{2}b_{--} \\
&  &  &  &  &  & 0 & a_{3}b_{--} \\
&  &  &  &  &  &  & 0
\end{array}
\right] ,  \label{A_c_3}
\end{equation}
%%\end{widetext}
\noindent where the vector $\boldsymbol{\xi} $ of $8$ basis states is
\begin{equation}
\boldsymbol{\xi} =\left[ +++ ;\,\,\, ++- ;\,\,\, +-+ ;\,\,\, ++- ;\,\,\, -++
;\,\,\, -+- ;\,\,\, --+ ;\,\,\, ---\right]^T ,  \label{basis_1}
\end{equation}
\noindent and the lower left portion of the symmetric matrix $A_C$ is
obtained by reflection with respect to the diagonal.

The driver $H_{E}$ is obtained by summation over all clauses. In doing this
summation, we take into account that now the bit indices $\left\{
i,j,k\right\} \in \left\{ 1,2,...n\right\} $ run through all $n$ bits,
whereas the indices $\left( \alpha ,\beta ,\gamma \right) \in \left\{
1,2,3\right\} $ characterizing the realization of $A_{C}$ still run through
the $3$ bits (since the same realization of $A_{C}$ is applied to all
triples of bits). The driver $H_{E}$ is given by
\begin{equation}
H_{E}=H_{E}^{\left( 1\right) }+H_{E}^{\left( 2\right) }+H_{E}^{\left(
3\right) },  \label{H_e_1}
\end{equation}
\noindent where $H_{E}^{\left( 1\right) },H_{E}^{\left( 2\right)
},H_{E}^{\left( 3\right) }$ correspond to the transitions involving one, two
and three bits, analogous to (\ref{A_c_1}). %%\begin{widetext}
\begin{eqnarray}
H_{E}^{\left( 1\right) } &=&\sum_{\alpha ,i,j,k}a_{\alpha }\ \sigma
_{i}^{x}\ \frac{1}{4}\sum_{s,s^{\prime }=\pm 1}b_{ss^{\prime }}\left( 1+s\
\sigma _{j}^{z}\right) \left( 1+s\ \sigma _{k}^{z}\right) ,  \label{H_e_2} \\
H_{E}^{\left( 2\right) } &=&\sum_{\alpha ,\beta ,i,j,k}a_{\alpha \beta }\
\left( \sigma _{i}^{+}\sigma _{j}^{+}+\sigma _{i}^{-}\sigma _{j}^{-}\right)
\ \frac{1}{2}\sum_{s=\pm 1}b_{s}\left( 1+s\ \sigma _{k}^{z}\right)   \notag
\\
&&\hspace{2.6in}+\widetilde{a}_{\alpha \beta }\ \left( \sigma _{i}^{+}\sigma
_{j}^{-}+\sigma _{i}^{-}\sigma _{j}^{+}\right) \ \frac{1}{2}\sum_{s=\pm 1}%
\widetilde{b}_{s}\left( 1+s\ \sigma _{k}^{z}\right) ,  \notag \\
H_{E}^{\left( 3\right) } &=&\sum_{i,j,k}B\left( \sigma _{i}^{+}\sigma
_{j}^{+}\sigma _{k}^{+}+\sigma _{i}^{-}\sigma _{j}^{-}\sigma _{k}^{-}\right)
+C\left( \sigma _{i}^{+}\sigma _{j}^{+}\sigma _{k}^{-}+\sigma _{i}^{-}\sigma
_{j}^{-}\sigma _{k}^{+}\right)   \notag \\
&&+\hspace{1.8in}D\left( \sigma _{i}^{-}\sigma _{j}^{+}\sigma
_{k}^{+}+\sigma _{i}^{+}\sigma _{j}^{-}\sigma _{k}^{-}\right) +E\left(
\sigma _{i}^{+}\sigma _{j}^{-}\sigma _{k}^{+}+\sigma _{i}^{-}\sigma
_{j}^{+}\sigma _{k}^{-}\right) .  \notag
\end{eqnarray}
%%\end{widetext}
\noindent One should note that the second term on the r.h.s. for $%
H_{E}^{\left( 2\right) }$ gives a contribution, which is diagonal in $S_{z}$
representation and therefore leads to the effective ''re-definition'' of the
cost function. Following the logic of \cite{Farhi:paths}, we disregard such
terms. Also, the commutation relations between the total spin components
give contributions $\thicksim 1/l$ to the effective potential and can be
neglected in the large-spin limit. Taking this into account, we obtain from (%
\ref{H_e_2}) in the large-spin limit %\\{\bf H$\_$e$\_$3a}
\begin{equation*}
H_{E}=\left( \frac{n}{2}\right) ^{3}\left( \gamma _{1}n_{x}+\gamma
_{2}n_{x}^{2}+\gamma _{3}n_{x}^{3}+\gamma _{4}n_{x}n_{z}+\gamma
_{5}n_{x}n_{z}^{2}+\gamma _{6}n_{x}^{2}n_{z}\right) ,
\end{equation*}
where $n_{a}=S_{a}/l$ is a dimensionless spin projection on $a$-axis and the
coefficients $\left\{ \gamma _{k}\right\} $ are given by
%\\{\bf gammas$\_$1}
%%\begin{widetext}
\begin{eqnarray}
\gamma _{1} &=&\frac{1}{6}\left( \sum_{\alpha }a_{\alpha }\right) \left(
\sum_{s,s^{\prime }}b_{ss^{\prime }}\right) +\frac{1}{3}\left(
C+D+E-3B\right) ,  \notag \\
\gamma _{2} &=&\frac{2}{3}\left( \sum_{\alpha ,\beta }a_{\alpha \beta
}\right) \left( b_{+}+b_{-}\right) ,  \notag \\
\gamma _{3} &=&\frac{1}{3}B,  \notag \\
\gamma _{4} &=&\frac{2}{3}\left( \sum_{\alpha }a_{\alpha }\right) \left(
b_{++}-b_{--}\right) ,  \label{gammas_1} \\
\gamma _{5} &=&\frac{1}{3}\left( \sum_{\alpha }a_{\alpha }\right) \left(
b_{++}+b_{--}-b_{-+}-b_{+-}\right) -\frac{1}{3}\left( C+D+E-3B\right) ,
\notag \\
\gamma _{6} &=&\frac{2}{3}\left( \sum_{\alpha ,\beta }a_{\alpha \beta
}\right) \left( b_{+}-b_{-}\right) .  \notag
\end{eqnarray}
%%\end{widetext}
\noindent In particular, the deterministic driver considered in \cite
{Farhi:paths} corresponds to $b_{-+}=b_{+-}=\left\{ b_{s}\right\} =\left\{
a_{\alpha \beta }\right\} =B=C=D=E=0$ , $a_{\alpha }=1$ and $%
b_{++}=-b_{--}=-2.$ It follows from (\ref{gammas_1}) that in this case, the
only non-zero coefficient in (\ref{gammas_1}) is $\gamma _{4}=-8$. This
corresponds to $H_{E}=-4nS_{x}S_{z}$, which is equivalent to $%
H_{E}=-2n\left( S_{x}S_{z}+S_{z}S_{x}\right) $ in the large-spin limit
according to the above discussion. %%\end{widetext}

\section{\label{bifurc} Bifurcation point analysis}

Taking into account only the $\gamma _{4}$ term in $H_{E}$ and expanding up
to the 4th order, we obtain the conditions $U^{\prime }=U^{\prime \prime
}=U^{\prime \prime \prime }=0$ for the $A_{3}$ bifurcation point $\left\{
\tau _{c},\gamma _{c},x\right\} $ in the form \cite{GILMORE1} (cf. (\ref
{condA3})) %%\begin{widetext}
\begin{eqnarray}
\tau _{c}\left( \beta _{1}+2\beta _{2}x+3\beta _{3}x^{2}\right)  &=&-\left(
1-\tau _{c}\right) \left[ 2x-x^{3}+\gamma _{c}\tau _{c}\left( 1-\frac{3}{2}%
x^{2}\right) \right] ,  \notag \\
\tau _{c}\left( 2\beta _{2}+6\beta _{3}x\right)  &=&-\left( 1-\tau
_{c}\right) \left[ 2-3x^{2}+\gamma _{c}\tau _{c}\left( -3x\right) \right] ,
\label{bif_2} \\
6\tau _{c}\beta _{3} &=&\left( 1-\tau _{c}\right) \left( 6x+3\gamma _{c}\tau
_{c}\right) ,  \notag
\end{eqnarray}
%%\end{widetext}
\noindent which have to be solved for $\gamma _{c}$ , $\tau _{c}$ and $x$
for the given $\left\{ \beta _{k}\right\} $. Solving for $x$, we obtain
condition (\ref{bifurcation_1}) in the text.


\begin{thebibliography}{99}
\bibitem{Farhi:paths}  E. Farhi, J. Goldstone, and S. Gutmann, ``Quantum
Adiabatic Evolution Algorithms with Different Paths'',
arXiv:quant-ph/0208135.


\bibitem{Farhi}  E. Farhi, J. Goldstone, S. Gutmann, and M. Sipser,
``Quantum Computation by Adiabatic Evolution'', arXiv:quant-ph/0001106,
(2002).

\bibitem{Farhi:Sc}  E. Farhi, J. Goldstone, S. Gutmann, J. Lapan, A.
Lundgren, and D. Preda, \lq\lq A quantum adiabatic evolution algorithm
applied to random instances of NP-complete problem'', Science, \textbf{292},
472 (2001).

\bibitem{Farhi:annealing}  E. Farhi, J. Goldstone, S. Gutmann, .''Quantum
Adiabatic Evolution Algorithms versus Simulated Annealing'',
arXiv:quant-ph/0201031 v1, (2002).

\bibitem{FarhiSat}  E. Farhi, J. Goldstone, and S. Gutmann, \lq\lq A
numerical study of the performance of a quantum adiabatic evolution
algorithm for Satisfiability'', arXiv:quant-ph/0007071.

\bibitem{FarhiCli}  A. M. Childs, E. Farhi, J. Goldstone, and S. Gutmann,
``Finding cliques by quantum adiabatic evolution'', arXiv:quant-ph/0012104.


\bibitem{Vazirani:02}  W. Van Dam, M. Mosca, U. Vazirani, "How Powerful is
adiabatic Quantum Computation?", arXiv:quant-ph/0206003.

\bibitem{Bennett}  C. Bennett, E. Bernstein, G. Brassard, and U.
Vazirani,''Strengths and weaknesses of quantum computing'', SIAM Journal of
Computing, \textbf{26}, pp. 1510-1523 (1997); arXiv:quant-ph/9701001

\bibitem{Vazirani:01}  W. Van Dam, M. Mosca, U. Vazirani, ''How Powerful is
Adiabatic Quantum Computation?'', FOCS 2001.

\bibitem{Vazirani:talk}  U. Vazirani, ''Quantum Adiabatic algorithms'', talk
on ITP Conference on Quantum Information, (UC Berkley, December, 2001),
http://online.itp.ucsb.edu/ on-line/qinfo$\_$c01

\bibitem{Wernsdorfer}  W. Wernsdorfer, R. Sessoli, \lq\lq Quantum phase
interference and parity effects in magnetic molecular clusters'', Science,
\textbf{284}, p.133 (1999).

\bibitem{Brooke:Nature}  J. Brooke, T.F. Rosenbaum and G. Aeppil, \lq\lq
Tunable quantum tunnelling of magnetic domain walls", Nature, \textbf{413},
p. 610 (2001).

\bibitem{Anderson}  Y. Fu and P.W. Anderson, \lq\lq Application of
statistical mechanics to NP-complete problems in combinatorial
optimization'', J. Phys. A: Math. Gen. \textbf{19}, 1605-1620 (1986).

\bibitem{Parizi}  M. Mezard, G. Parizi, and M.A. Virasoro, \emph{Spin Glass
Theory and Beyond}, (World Scientific, Singapore, 1987).

\bibitem{Monasson}  R. Monasson and R. Zecchina, ``Entropy of the
K-Satisfiability problem'', Phys. Rev. Lett., \textbf{76}, p.3881 (1996);
\emph{ibid}, ``Statistical mechanics of the random K-Satisfiability
problem'', Phys. rev. E \textbf{56}, p.1357 (1997).

\bibitem{SMEL1}  V. N. Smelyansky and U. V. Toussaint, ''Number Partitioning
via Quantum Adiabatic Computation'', arXiv:quant-ph/0202155.

\bibitem{adiabatic_theorem}  A. Messiah, \textit{Quantum Mechanics} (Wiley,
1976).


\bibitem{CHUD1}  E. M. Chudnovsky and D. A. Garanin, \lq\lq Quantum
tunnelling of Magnetization in small ferromagnetic particles'', Phys. Rev.
Lett., v.79, 4469 (1997).

\bibitem{GARG1}  A. Garg, ''Topologically Quenched Tunnel Splitting in Spin
Systems without Kramers' Degeneracy'', Europhys. Lett., v.22, 205 (1993).

\bibitem{GARG2}  M. Stone, K. Park, and A. Garg, ''The semiclassical
propagator for spin coherent states'', Journ. Math. Phys., v.41, 8025 (2000).

\bibitem{KLAUDER1}  J. R. Klauder, ``Path integrals and stationary-phase
approximation'', Phys. Rev. D \textbf{19}, p.2349 (1979).

\bibitem{LANDAU1}  L. D. Landau and E. M. Lifshitz, \emph{Quantum Mechanics}%
, (Pergammon, 1992).

\bibitem{ENZ1}  M. Enz and R. Schilling, ``Spin tunnelling in the
semiclassical limit'', J. Phys. C: Solid State Phys. \textbf{19}, 1765-1770
(1986).

\bibitem{GILMORE1}  R. Gilmore, \emph{Catastrophe Theory for Scientists and
Engineers}, (Wiley, 1981).

%\bibitem{Kadowaki}  T. Kadowaki and H. Nishimori, ``Quantum annealing in the
%transverse Ising model'', Phys. Rev. E \textbf{58}, p.5355 (1998).

\end{thebibliography}
\end{document}